*Subject Section*

# Elucidation of time-dependent systems biology cell response patterns with time course network enrichment

Christian Wiwie[1, *] Alexander Rauch[2], Anders Haakonsson[2], Inigo Barrio-Hernandez[2], Blagoy Blagoev[2], Susanne Mandrup[2], Richard Röttger[1], Jan Baumbach[1,3]

[1] Institute of Mathematics and Computer Science, University of Southern Denmark, Odense, Denmark

[2] Department of Biochemistry and Molecular Biology, University of Southern Denmark, Odense, Denmark

[3] Computational Systems Biology, Max Planck Institute for Informatics, Saarbrücken, Germany

*To whom correspondence should be addressed.

## Abstract

**Motivation:** Advances in OMICS technologies emerged both massive expression data sets and huge networks modelling the molecular interplay of genes, RNAs, proteins and metabolites. Network enrichment methods combine these two data types to extract subnetwork responses from case/control setups. However, no methods exist to integrate time series data with networks, thus preventing the identification of time-dependent systems biology responses.
**Results:** We close this gap with Time Course Network Enrichment (TiCoNE). It combines a new kind of human-augmented clustering with a novel approach to network enrichment. It finds temporal expression prototypes that are mapped to a network and investigated for enriched prototype pairs interacting more often than expected by chance. Such patterns of temporal subnetwork co-enrichment can be compared between different conditions. With TiCoNE, we identified the first distinguishing temporal systems biology profiles in time series gene expression data of human lung cells after infection with Influenza and Rhino virus.
**Availability:** TiCoNE is available online (https://ticone.compbio.sdu.dk) and as Cytoscape app in the Cytoscape App Store (http://apps.cytoscape.org/).
**Contact:** example@example.org
**Supplementary information:** Supplementary data are available at *Bioinformatics* online.

## 1 Introduction

Over the last decades, thanks to major advances and heavily decreasing costs in high-throughput experimental technology, we witnessed a sheer explosion in the amount and type of so-called OMICS data. The Gene Expression Omnibus (GEO) database, just to name an example, stores expression data for almost two million samples (Edgar, et al., 2002). Amongst them we find an increasing amount of time-series data sets measuring features (here: gene expression) over multiple time points and under different conditions.

At the same time, it becomes increasingly apparent that the measurements of activity of individual biological entities, such as transcription of genes or protein expression, are more helpful for understanding complex biological systems when analyzing them not only over time (Bar-Joseph, et al., 2012), but also in the context of their molecular interactions (Barabasi and Oltvai, 2004; Bracken, et al., 2016). The collection of high-quality interaction networks is one of the main goals of systems biology, and substantial improvements in technologies lead to a rapid growth of interactome databases (Lehne and Schlitt, 2009), such as BioGRID (Chatr-Aryamontri, et al., 2015) or I2D (Brown and Jurisica, 2005).

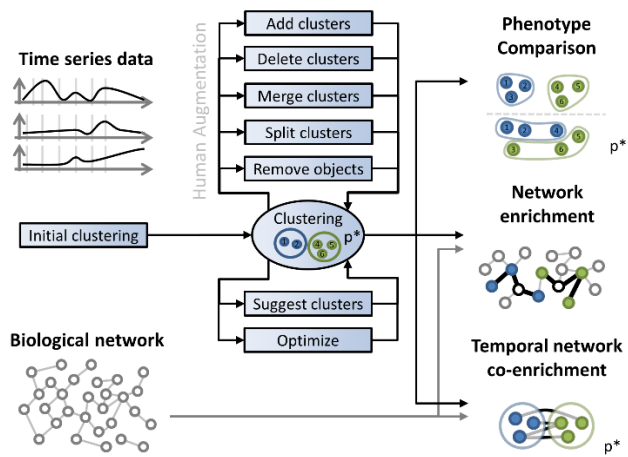

**Figure 1. Overview.** Time Course Network Enrichment (TiCoNE) is a novel unsupervised computational approach to enrich experimental time series data for overrepresented patterns by combining human-augmented clustering with methods for unraveling interacting systems in biomolecular networks enriched with temporal expression patterns. We developed tailored statistical models to assess enrichment significance at different steps of data analysis (p*).

Suitable approaches to analyze either of these data types separately exist: (1) Integrating computational approaches with time-series profiling have proven successful in discovering novel biological insights (Chylek, et al., 2014; Reddy, et al., 2016). Discovery of enriched temporal patterns in time-series data sets is usually based on time-behavior similarities and mostly identified using standard clustering methods (Warren Liao, 2005). Only few dedicated time-perspective centered tools have been developed: One of them is STEM (Ernst and Bar-Joseph, 2006), a clustering approach for time-series data. Other approaches, such as SAM (Tusher, et al., 2001) or EDGE (Leek, et al., 2006), identify temporally differentially expressed genes. (2) In addition, *de novo* network enrichment methods can find sub-networks enriched with object (e.g. genes) that are significantly differentially expressed between two different experimental conditions. Examples for such approaches are DEGAS (Ulitsky, et al., 2008) and KeyPathwayMiner (Alcaraz, et al., 2014).

Some methods exist that bridge between the two areas to a limited degree. VANTED (Rohn, et al., 2012) is a systems biology framework, tackling a larger set of common bioinformatics tasks. This includes data visualization, and basic cluster analysis using Self Organizing Maps. It also allows to map cluster objects to a network and provides network visualizations. However, it does not provide more sophisticated network analysis functionality, such as de-novo network enrichment. NeAT (Brohée, et al., 2008) is a collection of programs that allow for basic network integration of cluster results. However, it does not provide integrated clustering functionality but only allows for the import of clustering results derived with external tools. NEAT (Signorelli, et al., 2016) allows for the connectivity analysis of clusters on a network level. However, it does not provide any clustering facilities and expects pre-calculated clusters as inputs.

Overall, none of the existing approaches allow for the integrated analysis of differential time series expression clusters, with molecular interaction networks to illuminate systems biological response patterns of temporal resolution (see Supplementary Table 1 and 2).

Apart from the necessity to provide integrated approaches for time-series data with interaction networks, we firmly believe, that bioinformatics tools can only optimally support researchers in carrying out their analyses, if results are presented in an easily comprehendible way, and if researchers can incorporate their expert knowledge into the computational process.

Some approaches provide pre-defined visualizations for their results (see Supplementary Table 1 and 2), for instance NeAT (Brohée, et al., 2008), STEM (Ernst and Bar-Joseph, 2006), and VANTED (Rohn, et al., 2012). Clustrophile (Demiralp, 2016) also allows the user to interactively modify the two-dimensional data projection that is used to visualize the clustering result. However, almost none of the existing approaches expose interactive interfaces to the user and they derive their results in a purely automated manner. Few, so-called human-augmented clustering approaches exist, that allow the researcher to incorporate their knowledge into the clustering process. For instance, VISTA (Keke and Ling, 2003) allows for the interactive merging and splitting of clusters in the current clustering solution.

To fill these gaps, we have developed Time Course Network Enrichment (TiCoNE), a combination of a novel human-augmented time-series clustering method and a new kind of biological network enricher to identify putative biologically relevant regions in the network. Note that TiCoNE works with most kinds of biological entities (genes, proteins, RNAs, etc.) and most types of molecular measures acquirable for them, for instance, expression (transcriptomics, proteomics). As we focus on transcriptomics data later, and to increase readability, we simply refer to genes and gene expression in the following. We have summarized the most important features of TiCoNE in Supplementary Table 1.

## 2 Methods

The goal of any hard, partitional clustering method is to partition a given data set $X = \{O_1, ..., O_N\}$ with objects $O_i$ into a clustering $C(X) = \{c_1 ... c_k\}$, with clusters $c \subseteq X$, where $\bigcup_{c \in C(X)} c = X$ and $\bigcup_{c_p \neq c_q \in C(X)} (c_p \cap c_q) = \emptyset$. We write $C$ instead of $C(X)$, if the data set $X$ is clear from the context. We denote the cluster of an object $O \in X$ in clustering $C$ as $C(O)$. Since we focus on genes as objects in this work, we often refer to an object as a gene in the main text.

### 2.1 Input and Preprocessing

A time-series data set consists of time-dependent measurements for a collection of objects (here: genes), where $R \geq 1$ replicates are measured for each object. We formalize the objects $O \in X$ of a time-series data set as a collection of $R$ replicates $O = (O^1, ..., O^R)$, with each replicate being defined as a time series over $T$ time points: $O^r = (O^r_1, ..., O^r_T), 1 \leq r \leq R, 1 \leq t \leq T$. In the remainder, the term "object" includes all of its replicate time-series.

The preprocessing of raw data is a highly setting-specific task, and it should thus be performed by the user prior to clustering in TiCoNE. Nevertheless, we provide two basic data preprocessing operations:

- Remove baseline objects, i.e. $X^* = X \setminus \{O \in X | \exists \leq r \in 1...R: \overline{\sigma(O^r)} < \sigma^*\}$, where $\sigma^*$ is a user-given threshold for the average standard variance of objects to keep.
- Remove objects $O$ with low agreement between replicates: $X^* = X \setminus \{O \in X | A(O) < A^*\}$, where $A^*$ is a user-given threshold for the replicate agreement of objects to keep, and $A(O)$ the replicate agreement of object $O$ as defined in the next section.

## 2.2 Similarities

The objective of clustering is to group the genes in such a way, that the time courses in one cluster are more similar to each other than to the time courses of other clusters. We denote the similarity of two genes as $S(O_i, O_j) = 1/R^2 \cdot \sum_{r=1}^{R} \sum_{r=1}^{R} S(O_i^r, O_j^r)$.

We support two measures to assess the similarity $S(O_i, O_j)$ between time courses of two genes $O_i, O_j \in X$: (1) the (negative) Euclidean distance $S_E(O_i, O_j)$ and (2) the (scaled) Pearson correlation $S_P(O_i, O_j)$. We define the negative Euclidian distance of two vectors $x, y \in \mathbb{R}^T$ as $S_E(x, y) = -d_E(x, y) = -\sqrt{\sum_{t=1}^{T}(x_t - y_t)^2}$ and the scaled Pearson Correlation as $S_P(x, y) = (\rho(x, y) + 1)/2$.

The former assesses the similarity of the time courses according to the magnitudes, the latter according to their shape. We denote the case $S(O, O) = A(O)$ and call it the agreement of object $O$'s replicates. The similarity of an object $O$ to a prototype $P$ can be defined as $S(O, P) = 1/R \cdot \sum_{r=1}^{R} S(O^r, P)$.

## 2.3 Initial Clustering & Iterations

TiCoNE performs iterations $i \in \{1 \dots F\}$, where F denotes the last performed iteration. We denote the clustering in iteration $i$ as $C^i = \{c_1^i, \dots, c_{|C^i|}^i\}, 1 \leq i \leq F$, where we refer to $C^F$ as the final clustering. If a clustering $C$ is denoted without iteration, we assume that $C = C^F$. TiCoNE integrates the following clustering methods to find an initial clustering $C^1$ with a user-specified number of clusters $|C^1|$: CLARA (Kaufman and Rousseeuw, 2008), k-Means (MacQueen, 1967), PAMK (Kaufman and Rousseeuw, 2008), STEM (Ernst and Bar-Joseph, 2006), or Transitivity Clustering (Wittkop, et al., 2010). For all clusters $c \in C^1$ new prototypes $\langle c \rangle$ are calculated. $C^1$ can then be inspected and iteratively refined in later iterations resulting in clusterings $C^2, C^3, \dots, C^F$.

## 2.4 Prototypes & Clustering Optimization Iterations

TiCoNE follows a prototype-based approach (Borgelt, 2006; Tan, 2006) similar to k-means (MacQueen, 1967). With such approaches, each cluster $c \in C$ is represented by a prototype $\langle c \rangle$, which is a representation of objects (i.e. genes) $O \in c$. Usually, $\langle c \rangle$ is derived from all $O \in c$ using some aggregation function $f_A$. We define $\langle c \rangle = (f_A(c, 1), \dots, f_A(c, T))$ with $T$ being the number of time points. We provide three aggregation functions in TiCoNE, where $r \in \{1 \dots R\}$ and $t \in \{1 \dots T\}$:

- The **mean**: $f_A(c, t) = \overline{\{O_t^r | O \in c\}}$
- The **median**: $f_A(c, t) = median\{O_t^r | O \in c\}$
- The **most central object**: $f_A(c, t) = \{O_t^{*r} | O^* = argmax_{O \in c} \sum_{O_k \neq O \in c} S(O, O_k)\}$

At any stage during the clustering process, the user can let TiCoNE optimize the current clustering by either: (1) performing one optimization iteration or (2) optimization iterations until convergence. Assume, we are in iteration $i_s$ with clustering $C^{i_s}$. Execution of one optimization iteration then consists of performing the following two steps, where $O \in X$:

(1) Reassign genes to the most similar prototype, i.e. $C^{i_s+1}(O) = argmax_{c \in C^{i_s}} S(O, \langle c \rangle)$, where we break ties randomly.
(2) Recalculate prototypes $\langle c^{i_s+1} \rangle$ based on the new genes.
   a. Remove empty clusters, i.e. $C^{i_s+1} = C^{i_s} \setminus \{c \in C^{i_s} | c = \emptyset \} = \emptyset$ unless $c$ is marked as "keep".
   b. Clusters with identical prototypes are merged.

Iterations until convergence stops as soon as the following condition holds: $C^{F+1} = C^F \vee F \geq i_s + 100$, i.e. either the clustering was not changed in the iteration, or we performed 100 iterations.

## 2.5 Prototype Discretization

To give the user control over the tradeoff between sensitivity, i.e. detecting more and smaller clusters, and selectivity, i.e. finding less clusters of random noise, cluster prototypes are discretized and can only assume values $x \in N^- \cup N^+ \cup \{0\}$, where $N^-$ is a set of $n^- = |N^-|$ negative discretized values, and $N^+$ is a set of $n^+ = |N^+|$ positive discretized values. We define the sets of discretized negative and positive values as follows:

$$N^- = \left\{min, \frac{(n^- - 1)}{n^-} \cdot min, \frac{(n^- - 2)}{n^-} \cdot min, \dots, \frac{1}{n^-} \cdot min\right\}$$

$$N^+ = \left\{\frac{1}{n^+} \cdot max, \frac{2}{n^+} \cdot max, \dots, max\right\}$$

Where $min$ and $max$ denote the minimum and maximum input values observed for any gene at any time-point and replicate.

## 2.6 Human Augmentation Operations

With TiCoNE we offer a human augmented clustering approach, which implies that the user may (but does not have to) manually influence the clustering outcome. In contrast to fully automated clustering approaches, TiCoNE makes better use of field-specific knowledge that a fully automated procedure is not aware of. Human-augmented algorithms have been presented in several disciplines, such as robotics (Voyles, et al., 1999), linguistics (Chaiken and Foster, 2002), and also bioinformatics (Laczny, et al., 2015). In TiCoNE the user can modify a clustering in any of the following ways, where $F$ is the current iteration:

### 2.6.1. Merge clusters

A set of selected clusters $\hat{C} \subseteq C$ can be merged together into a new cluster $c^*$. The new cluster is defined as $c^* = \bigcup \hat{C}$ and the new clustering as $C^{F+1} = (C \setminus \hat{C}) \cup \{c^*\}$. The cluster's prototype $\langle c^* \rangle$ is initialized appropriately.

### 2.6.2. Split clusters

A cluster $c \in C$ can be split into a set of clusters $\hat{C}$ where $C^{F+1} = (C \setminus \{c\}) \cup \hat{C}$. This can be done in two different ways: (1) use objects $O_1 \neq O_2 \in c$ with $argmin_{O_1, O_2} S(O_1, O_2)$ to calculate prototypes $\langle \hat{c}_1 \rangle = \overline{O_1^r}$ and $\langle \hat{c}_2 \rangle = \overline{O_2^r}$ for two new clusters $\hat{C} = \{\hat{c}_1, \hat{c}_2\}$, or (2) objects $O \in c$ can be clustered into a user-specified number of smaller clusters using one of the available clustering methods.

### 2.6.3. Keep a prototype

A cluster $c$ can be marked as "keep", here denoted as $\vec{c}$. If a cluster has been marked as keep in iteration $F$ it holds that $\langle \overline{c^F} \rangle = \langle \overline{c^{F+1}} \rangle$ and $\overline{c^F} \in C^F \Leftrightarrow \overline{c^{F+1}} \in C^{F+1}$, even if $\overline{c^F} = \emptyset$. The "keep" flag can always be revoked in later iterations.

### 2.6.4. Add cluster with predefined prototype

A cluster $c$ with a user-specified prototype $\langle c \rangle$ can be added to the clustering. Initially it holds that $c = \emptyset$. To prevent immediate removal of $c$ from $C$ we set it to keep, i.e. $c = \vec{c}$.

### 2.6.5. Delete cluster or its objects

The user can delete a cluster $c \in C$ and (1) delete its genes from the data set and its prototype from the clustering. (2) Delete its prototype, but keep its genes. (3) Delete its genes but keep its prototype or (4) delete its least-fitting genes, i.e. $c^{F+1} = c \setminus \{O \in c | S(O, \langle c \rangle) < S^*\}$, where $S^*$ is a user-defined similarity threshold.

### 2.6.6. Handle least-fitting objects

The user has the possibility to remove genes from the clustering that are least similar to their respective prototypes, i.e. it holds that $\forall c \in C: c^{F+1} = c \setminus \{O \in c | S(O, \langle c \rangle) < S^*\}$, where $S^*$ is a user-defined similarity threshold. The deleted genes can either be (1) completely removed from the clustering or (2) clustered into a user-specified number of new clusters.

## 2.7 Clustering History & Reproducibility

Throughout the clustering process TiCoNE keeps a history of all performed steps. This includes automatically performed iterations, as well as human augmentations. At any time, the user can inspect clusterings $C_1, \ldots, C_F$ and, if desirable, revert the current clustering to a previous stage. The history can be exported for publication documenting the clustering process.

## 2.8 Network Enrichment of Clusters

We provide two ways of identifying subnetworks $\hat{G} = (\hat{V}, \hat{E}) \subseteq G = (V, E)$ enriched in genes of a selected set of clusters $\hat{C} \subseteq C$: (1) Extracting node-induced subnetworks with $\hat{V} = \bigcup \hat{C}$ and $\hat{E} = \{\hat{V} \times \hat{V} \in E\}$. Such networks are algorithmically easy to compute. However, this approach will fail to find connected networks if the genes of a cluster are not directly linked in the network; (2) In a biomedical setting it is reasonable to assume, that not all functionally related objects show a very similar time behavior and end up in the same cluster. Thus, we may want allow for a certain number of exception nodes that are not in the selected clusters but connect other objects that are. We use KeyPathwayMiner (Alcaraz, et al., 2014) to perform this task. It extracts a maximal connected subnetwork consisting only of genes from the selected clusters but a user-given number of exceptions.

## 2.9 Statistics

TiCoNE makes use of different fitness scores in different contexts to calculate empirical p-values, for example for clusters. If multiple fitness scores should be used to derive an overall p-value, a p-value is calculated for each fitness score independently, which are then combined into an overall p-value (see below).

For an observed instance (e.g. cluster), the p-value for one specific fitness score is calculated by comparing the original fitness against fitness scores of instances from a background distribution. The p-value is defined as the number of times, we observe fitness scores at least as extreme as the original one.

Let $FS$ be a fitness score, let $Z^* = (z_1^*, \ldots, z_Q^*)$ be the permuted instances, and let $\mathbf{1}(b)$ be the indicator function, defined as follows:

$$\mathbf{1}(b) = \begin{cases} 1, & \text{if } b \\ 0, & \text{otherwise} \end{cases}, b \in \{\text{true, false}\}$$

In a one-sided setting, the p-value for $FS$ and $z$ is then defined as:

$$p(z; FS) = \frac{1}{Q} \cdot \sum_{z^* \in Z^*} \mathbf{1}(FS(z^*) \geq FS(z))$$

And in a two-sided setting as:

$$p(z; FS) = \frac{1}{Q} \cdot \left( 2 * \min\left( \sum_{z^* \in Z^*} \mathbf{1}(FS(z^*) > FS(z)), \sum_{z^* \in Z^*} \mathbf{1}(FS(z^*) < FS(z)) \right) + \sum_{z^* \in Z^*} \mathbf{1}(FS(z^*) = FS(z)) \right)$$

We combine p-values of multiple fitness scores $FS_1, \ldots, FS_R$ for an instance $z$ by using a function $CO$: $p(z) = CO(p(z; FS_1), \ldots, p(z; FS_R))$. TiCoNE combines p-values with $CO = \prod \max(p(z; FS_i), PSEUDO)$, where PSEUDO refers to a pseudo count constant (default: 0.0001).

### 2.9.1. Generating background distributions

TiCoNE provides multiple ways of generating background distributions for p-value calculations.

1. Permuting Clusterings: We provide several ways of creating random clusterings: (PC1) Take a given clustering, shuffle the values of the cluster prototypes over time, reassign objects to shuffled prototypes. (PC2) Generate $k$ random prototypes and assign objects to those. (PC3) Take a given clustering, and randomly assign objects to the cluster prototypes while maintaining the cluster sizes. (PC4) Take a given clustering and randomly assign objects to the cluster prototypes while trying to maintain the sum of the node-degrees in each clusters, given the network. We sort all objects according to the node degrees in the network (descending). We then greedily assign the objects from top to bottom to random cluster, but only those that have not reached the sum of degrees of the original cluster. Note that this method yields comparable p-values to methods permuting the network using edge cross-over but with substantially reduced computing time.

2. Permuting Time-Series Data: We employ two approaches to create a permuted version $X^*$ of a given time-series data set $X$: (PD1) Object-wise: We shuffle the values of an object across time points, i.e. we reassign each value once to a random time-point without replacement. Each gene is permuted independently, and thus it keeps the same distribution of values. If a gene has several replicates, they are shuffled in the same order. (PD2) Globally: Values are shuffled across all objects at the same time without replacement. Here the value distributions of the gene can change. Several replicates of the same gene are shuffled in the same order. In both approaches, each gene and time point is shuffled exactly once.

3. Permuting Networks: We create a permuted version $G^* = (V^*, E^*)$ of a given network $G = (V, E)$ by performing edge crossovers (also called edge swaps). Let an edge be defined as $e = (v_1, v_2) \in E$, where $v_1, v_2 \in V$. We call $v_1$ the source of $e$ and $v_2$ the target of $e$. Given two randomly chosen edges $e_1 = (v_1, v_2), e_2 = (v_3, v_4) \in E$, we swap their targets and thus $E = E \setminus \{e_1, e_2\} \cup \{(v_1, v_4), (v_3, v_2)\}$. In case of undirected networks, source and target of an edge needs to be treated interchangeably when swapping. In order to ensure that the permuted network is sufficiently randomized we swap $|E|/2 \cdot c$ times, where $c = \ln(10^7)$ (Ray, et al., 2014).

## 2.10 Significance of Clusters

TiCoNE calculates empirical p-values for clusters based on the permutation scheme outlined above. The fitness scores for cluster p-values are the number of objects, $FS_1 = |c|$ and the average object-prototype similarity, $FS_2 = \overline{s(c)} = 1/|c| \cdot \sum_{O \in c} S(O, \langle c \rangle)$. Thus, cluster p-values approximate the probability of observing a cluster $c^*$ at random with $|c^*| \geq |c|$ and $\overline{s(c^*)} \geq \overline{s(c)}$. In our example study we use permutation method PC2 to generate random clusterings.

### 2.11 Time Course Network Co-Enrichment

TiCoNE allows for analyzing which pairs of clusters are connected more (less) strongly than expected in a given network (temporal cross-talk enrichment). We use our permutation scheme to calculate empirical p-values (either one-sided or two-sided) as well as the expected number of edges between the elements of two clusters in the network. We use the fitness score $FS = E_\rightarrow(c_1, c_2)$ in the case of directed networks and $FS = E_\leftrightarrow(c_1, c_2)$ for undirected networks, where $E_\rightarrow(c_1, c_2)$ is the directed number of edges from cluster $c_1$ to cluster $c_2$ and $E_\leftrightarrow(c_1, c_2)$ is the undirected number of edges between cluster $c_1$ to cluster $c_2$.

We calculate the expected number of edges between clusters $EE_\leftrightarrow(c_1, c_2)$ (undirected) and $EE_\rightarrow(c_1, c_2)$ (directed) as the average numbers observed across all permutations. We then calculate a log-odds-ratio for two clusters $c_1, c_2 \in C$ as $OR(c_1, c_2) = \log_2(E_\rightarrow(c_1, c_2)/EE_\rightarrow(c_1, c_2))$. Using this formula TiCoNE compares the expected number of edges with the observed number of edges for each cluster pair. In our example study we utilized method PC4 to generate a background distribution for co-enrichment p-values.

### 2.12 Phenotype Comparison

If biological entities have been analyzed under more than one experimental condition and are given as two data sets $X$ and $Y$, the first step is to cluster the objects for each setting separately and then compare the resulting clusterings $C(X)$ and $C(Y)$ for differentially clustered object (e.g. gene) sets. TiCoNE can identify such objects by comparing each cluster $c_1 \in C(X)$ with each cluster $c_2 \in C(Y)$. It then presents these clusters with their normalized number of common objects $NCO(c_1, c_2) = CO(c_1, c_j)/\min(|c_1|, |c_j|)$ and the similarity of their prototypes $S(\langle c_1 \rangle, \langle c_2 \rangle)$.

We assume that a cluster pair $(c_1, c_2)$ from two different phenotypes is interesting, if $c_1$ and $c_2$ have more objects in common than expected by chance for a pair of random clusters with a prototype similarity at most this high. We calculate p-values for each pair of clusters $(c_1, c_2) \in C(X) \times C(Y)$ using our permutation scheme and the fitness score $FS = NCO(c_1, c_2) * S(\langle c_1 \rangle, \langle c_2 \rangle)$. For our example study we generate random clusterings using permutation method PC2.

#### 2.12.1. Network-Enrichment of Differential Phenotype

Once phenotypes have been compared, TiCoNE can identify subnetworks that are enriched in genes with a differential time behavior. Again, we employ the KeyPathwayMiner approach (see above) to identify all maximal connected subnetworks where all genes but a certain number of exceptions show the selected differential behavior. In our example study, we set the number of allowed exceptions to 0.

### 2.13 Synthetic Data

We generated a synthetic data set $X$ for illustration purposes, that consists of (1) genes belonging to three clusters with a distinct predefined time behavior (see Supplementary Figure 1) and (2) background genes with random time behavior. The time series signals of the cluster prototypes lie in the range of $[-3,3]$. For each cluster we generated 80 objects, each with $R = 2$ time-series (replicates) and $T = 5$ time points. We required $S_P(O, \langle C(O) \rangle) \geq 0.6$ and $\forall_{t \in 1..T, r \in 1..R}: |O_t^r - \langle C(O) \rangle_t| \leq 1.8$ for all objects (i.e. genes) $O \in X$. Thus, the value range of the objects is $[-4.8, 4.8]$. For the 120 background objects we generated time series with a random behavior, in the same value ranges and with the same number of samples.

The network describes interactions of the 240 genes of the three clusters and of the 120 randomly behaving background genes. To demonstrate how our co-enrichment feature works, we generated two clusters with similar node degree distributions, but converse connectivity to genes of the other clusters. See Supplementary Figure 1 for an illustration and its legend for details regarding the concrete patterns.

#### 2.13.1. Clustering

We clustered the data set into 10 clusters using the CLARA clustering method, 15 negative and positive discretization steps, the Pearson correlation and cluster aggregation function $f_A = \mu$.

### 2.14 Influenza and Rhino Virus Data

We applied TiCoNE to human gene expression data measured with an Affymetrix Human Genome U219 Array containing expression levels for 49,386 probes over 10 time points (baseline and at 2hrs, 4hrs, 8hrs, 12hrs, 24hrs, 36hrs, 48hrs, 60hrs, 72hrs after infection) under three different experimental conditions: BEAS-2B lung cells have been infected with (1) Influenza virus, (2) Rhino virus, or (3) co-infected with both. The data set contained five biological replicates for all time points except the first one, for which it contained six. See ref. (Kim, et al., 2015) for more details. It is an ideal scenario for TiCoNE's phenotype comparison feature. Here, we focus on comparing (1) Influenza vs. (2) Rhino virus infection.

We mapped the probe set IDs of the data set to Entrez gene IDs. For genes with multiple probesets measured we kept only the one with the highest variance over time. The input values have already been log2-transformed such that we only normalized them against the control. Furthermore, we removed genes from the data sets, which were not present in the I2D network (see below).

#### 2.14.1. Clustering

We clustered both, the expression data of the Influenza and the Rhino virus infection experiments with CLARA and 20 negative/positive discretization steps into 500 initial clusters. We used the Pearson correlation as similarity function, cluster aggregation function $f_A = \mu$, and removed genes not present in the I2D network (see below). We then ran automatic iterations until convergence.

### 2.15 Biological Network

We extracted the human interactome from the I2D protein-protein interaction (PPI) database (Brown and Jurisica, 2005), which integrates various other databases of known, experimental and predicted PPIs. We mapped Uniprot IDs to entrez gene IDs to be able to integrate the network with our example time-series data. After removing duplicate interactions between the same pair of genes the resulting network contains 199,025 interactions and 15,161 nodes (genes). Note that our TiCoNE approach works with any kind of graph loaded into Cytoscape or to the TiCoNE online platform. The use case determines the most appropriate network. Here, we aimed for finding human response protein complex formation candidates; hence our choice of using the I2D interactome.

## 3 Results

Figure 1 illustrates the structure and unique capabilities of Time Course Network Enrichment: 1) One may find subnetworks significantly enriched with genes of a similar temporal expression behavior. 2) The same can be applied to several such clusters of genes, i.e. multi-temporal network expression enrichment. 3) Furthermore, the crosstalk between genes over time can be analyzed on a systems level by computing the likelihood of observing more (or less) network interactions between pairs of time patterns by chances; a procedure we call network co-enrichment analysis. 4) The TiCoNE software computes empirical p-values and cross-talk graphs. 5) Finally, one may analyze multiple condition-specific time series experiments to identify overrepresented temporal patterns responding differentially to conditions in multiple experiments.

To illustrate Time Course Network Enrichment we first utilize a synthetically generated data set, which we use for illustrative purposes only (Supplementary Figure 1, also to be found at the project web site). Afterwards, we will demonstrate its full power by analyzing whole-genome time series transcriptomics data of lung cells after infection with Influenza or Rhino virus together with the human protein interactome. This led to the identification of interesting overrepresented time-series patterns, for which we analyzed the interplay in the network. Finally, we compared the systems biology response between the two infection types.

The TiCoNE software is publicly available as a Cytoscape app (Shannon, et al., 2003), downloadable from the Cytoscape app store, and a feature-reduced web-version hosted at https://ticone.compbio.sdu.dk together with tutorials and screencasts.

### 3.1 Illustration on Synthetic Data

We illustrate TiCoNE's capabilities on artificial time course data and a corresponding synthetic network: We generated a time-series data set containing three object clusters and a set of random objects, RND. The three clusters exhibit a pre-specified time-behavior (prototypes P1-P3, see Supplementary Figure 1). Correspondingly, we create an interaction network of all objects with pre-specified connectivities between the clusters and random objects (Supplementary Figure 1).

Initially, we clustered the data set into k=10 clusters, pretending not to know the real number of clusters. See Methods section for a detailed description of the data set and the used parameters. In the resulting clustering, the three hidden clusters are each split up into several highly similar subclusters. This is expected, since k=10>3. We used TiCoNE's human-augmented merge function (see Methods) to reduce redundancy, and ended up with eight final clusters (Supplementary Figure 2a and Supplementary Table 1). For each of the clusters, TiCoNE computed a consensus pattern that we call prototype (Methods).

Three of the identified clusters show a high object-prototype Pearson correlation as well as a significant p-value. The other five clusters show lower similarities as well as less to non-significant p-values (see Supplementary Table 1). A manual investigation of the prototypes confirms that clusters 5, 9, and 11 correspond to the three gene clusters that we have hidden in the network and aimed to recover. However, several background objects have been clustered into our three foreground clusters: Cluster 5 (7), cluster 9 (11), and cluster 11 (14). Also, 4 objects of cluster 9 are misplaced into random clusters.

Instead of trying to further reduce this number using the time course data alone we employed TiCoNE's network functionality. Visual inspection of the clusters on the network reveals three distinct groups (corresponding to our foreground clusters) and a large sets of obvious outlier objects. We used TiCoNE's time course network co-enrichment analysis (Supplementary Figure 2b) to recover the network-connectivities of clusters we hid in the network (see Methods): Cluster pairs (5, 5), (9, 9), (11, 11), (5, 9), (11, 1), (11, 4), (11, 8), (11, 10), and (11, 12) to be highly significantly (p=0.0) enriched with log-odds ratios OR>0.38 (observed compared to expected number of edges), while the cluster pairs (5, 11), (9, 11), (9, 10), (9, 12), (9, 8), (9, 4), (5, 10), (5, 1), (5, 12), (5, 8), (5, 4), are highly significantly (p=0.0) depleted with odds ratios OR≤-0.217 and cluster pair (1,9) is significantly (p≤0.05) depleted with odds ratio OR=-0.138.

### 3.2 Differential Interactome Time-Response to Influenza and Rhino Virus Infection

After illustrating that Time Course Network Enrichment can retrieve artificially hidden temporal expression patterns in gene sets and networks, we apply it to a real data set in the following. We utilized human whole-genome gene expression data of lung cells after infection with Influenza virus and Rhino virus 22. The data set contains expression levels for 49,386 probes over ten time points (baseline and at 2hrs, 4hrs, 8hrs, 12hrs, 24hrs, 36hrs, 48hrs, 60hrs, 72hrs after infection) with five replicates for each time point. We concentrated on two experimental conditions: BEAS-2B lung cells that have been infected (1) with Influenza virus and (2) with Rhino virus. See Methods for more details.

We clustered both, the expression data of the Influenza infection (shortly "Influenza data" in the following) and the Rhino virus infection experiments (shortly "Rhino data") into 500 clusters. Since our method iteratively merges clusters with equivalent prototypes we ended up with 173 Influenza clusters (88 with p≤ 0.05) and 225 Rhino clusters (71 with p≤ 0.05) (see Supplementary Figures 3a and 3b for selections of significant clusters). All clusters have a time course behavior distinct from each other and most prototypes peak multiple times into different directions.

For each data set we performed a time course network co-enrichment analysis. When comparing the distributions of undirected edge counts between cluster pairs (see Supplementary Figure 7), the Influenza clusters (i.e. clusters in the "Influenza data") represent gene groups that are more tightly connected in the network (more undirected edges) than the Rhino clusters (i.e. clusters in the "Rhino data"). For both data sets, log-odds scores are slightly left skewed normally distributed, while overall cluster pairs of the Rhino data tend to be more depleted in edges than Influenza cluster pairs. We arranged clusters for both data sets according to their most significant connectivities (p≤ 0.02) on a timeline (see Supplementary Figures 4 and 5). Generally, Influenza clusters with significant connectivities (either depleted or enriched) form a large connected complex along the timeline. In contrast, such significant Rhino cluster pairs seem to be more fragmented and form multiple separated cluster complexes. Out of the 61 significant Influenza data cluster pairs, we see more often an enriched of the number of edges (39/61) than for the 34 significant Rhino data cluster pairs (19/34).

Finally, we use TiCoNE's phenotype comparison functionality to search for gene groups behaving temporally different under the two conditions. We identified 30 cluster pairs that show a highly significant ($p \leq 0.01$) enrichment in the number of common genes when compared to randomly generated clusters (see Methods). When enriching for the three cluster pairs with the smallest p-values ($p \leq 0.001$) in the interactome network (see Methods), we find a connected component consisting of 50 genes and 101 interactions (see Figure 2). This complex contains 30 genes (60%) that have been annotated as relevant in immune responses to Influenza virus. Of these 30 genes, 27 have been discussed in the literature: DHX58 (also called LGP2) (Malur, et al., 2012; Si-Tahar, et al., 2014), EIF2AK2 (Carter, 2009; Kumar, et al., 2017), GBP1 (Zhu, et al., 2013; Zhu, et al., 2013), HERC5 (Dastur, et al., 2006; Tang, et al., 2010), IFIH1 (Kobasa, et al., 2007; Neumann, et al., 2009), IFIT1 (Pichlmair, et al., 2011), IFIT2, IFIT3, IFIT5 (Diamond and Farzan, 2013), IRF1 (Iwasaki and Pillai, 2014), IRF2 (Shapira, et al., 2009), IRF7 (Ciancanelli, et al., 2015), IRF9 (Kim, et al., 2015), ISG15 (Morales, et al., 2015), ISG20 (Degols, et al., 2007; Espert, et al., 2003), LGALS3BP (Wilk, et al., 2015), MX1 (Shin, et al., 2015; Tumpey, et al., 2007), MYD88 (Seo, et al., 2010), NLRC5(Ranjan, et al., 2015), PLSCR1 (Talukder, et al., 2012), PML (Gupta, et al., 2014; Iki, et al., 2005), RSAD2 (Gupta, et al., 2014; Iwasaki and Pillai, 2014), SP100 (Chelbi-Alix, et al., 1998), STAT1 (García-Sastre, et al., 1998), TLR3 (Le Goffic, et al., 2007), TNFSF10 (Jin, et al., 2014) and TRIM21 (Yang, et al., 2009). Some are also annotated with a differential behavior between Rhino and Influenza viral infection, for example IRF9 (Kim, et al., 2015), and STAT1 (Culley, et al., 2006; Ramirez, et al., 2014; Tyner, et al., 2005). In addition, 15 out of the 50 genes are contained in the KEGG pathway for Influenza A infection (hsa05164) (Kanehisa and Goto, 2000; Kanehisa, et al., 2016).

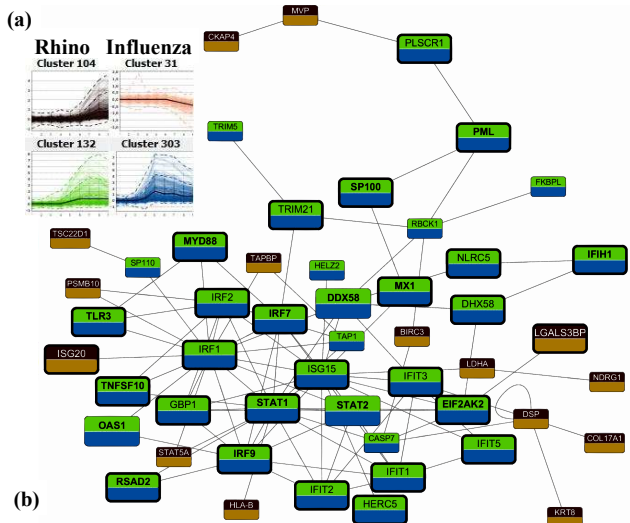

**Figure 2**. **Temporal host gene expression response comparison between Rhino virus and Influenza virus response patterns. (a)** Largest connected subnetwork enriched with genes of the two most significant cluster pairs (size: 50 genes). Node colors correspond to the genes' respective cluster in the Rhino virus (upper half) and Influenza virus data sets (lower half). Larger nodes correspond to 30 known Influenza-related genes, where we consider two kinds of annotations: Bordered nodes correspond to genes linked to Influenza in the literature; genes with bold font are contained in the Influenza A KEGG pathway. **(b)** The two most different pairs of temporal gene cluster pairs of Rhino virus vs. Influenza virus infection response.

When repeating the above procedure for all of the 30 significant cluster pairs ($p \leq 0.01$) we find a larger connected component consisting of 177 genes and 342 interactions (see Supplementary Figure 6). It contains additional relevant genes associated with Influenza virus infection, such as IFNB1 (Hillesheim, et al., 2014), IFI6 (Marazzi, et al., 2012), CASP1 (Huang, et al., 2013; Kim, et al., 2015), ICAM1 (Othumpangat, et al., 2016), IFI27 (Ioannidis, et al., 2012), CXCL11 (Ramirez, et al., 2014), and CXCL10 (Ramirez, et al., 2014).

## 4 Discussion

We developed Time Course Network Enrichment analysis, the first method to seamlessly identify regions in biomolecular networks that are enriched in genes (or proteins, metabolites) with significantly overrepresented time-series behavior or time-series co-expression behavior. Our TiCoNE implementation offers a human-augmented cluster optimization strategy and allows the user to iteratively refine the clustering automatically or visually by applying operations such as adding, deleting, merging or splitting clusters. Our approach can compare different phenotypes and identify differentially behaving network complexes. TiCoNE computes empirical p-values for clusters, phenotype comparisons and network co-enrichment based on different kinds of permutation tests (see Methods).

We illustrated the principle of Time Course Network Enrichment on synthetically generated data. The power of TiCoNE was then demonstrated by processing time-series gene expression data of human host lung cells, infected with Rhino virus or Influenza. Here, we use our co-enrichment analysis approach to construct complexes of clusters along a timeline that are biologically meaningful and may explain the systemic unraveling of host immune response to Influenza virus infection. We find specific properties of co-enrichments of Influenza and Rhino data clusters, which may help explain the large difference in severity of the two viruses on a systems biology level.

Additionally, we discovered de novo groups of genes that behave consistently but show temporally different behavior under the two conditions. By integrating these candidates with the human protein-protein interaction network we discovered a complex of 50 genes, out of which 30 have been previously associated with Influenza virus infection. We consequently hypothesize that the several of the remaining genes are promising candidates of also being involved in the Influenza-specific response of the human immune system.

In conclusion, with Time Course Network Enrichment we close an important gap in systems biology research by providing the first method to integrate time series data directly with molecular biological networks in a sophisticated manner. In addition, it can compare identified temporal systems biology patterns of cellular responses between two different conditions or phenotypes. It also implements a novel human-augmented approach to clustering time series data into patterns. Finally, note that TiCoNE cannot just handle time-resolved data, but in principle is directly applicable as well to multi-condition (instead of multi-time point) expression data. It is publicly available in the Cytoscape app store and can be applied to any kind of expression data and molecular network loaded into Cytoscape. At the project web site (ticone.compbio.sdu.dk) we also offer access to a feature-reduced version of TiCoNE as well as access to example data and screen cast videos describing the TiCoNE usage click by click.

## Funding

*Conflict of Interest:* none declared.

## Supplementary Table 1: Main Features of TiCoNE

Descriptions of the most important tasks and corresponding sub-tasks that can be carried out with TiCoNE. Tasks are supported in the TiCoNE Cytoscape App and the TiCoNE WEB version as indicated.

| Task | Sub-Task | Description | Supported in | |
|---|---|---|---|---|
| | | | *TiCoNE Cytoscape App* | *TiCoNE WEB* |
| *Data Handling* | Feature Scaling, feature selection/feature weights | Weights can be assigned to data features, to account for differences between them, for instance in noise levels or accuracies. | ✔ | ✗ |
| | Removing objects with low agreement between replicates | Objects that exhibit a low reproducibility across replicates can be removed from the data set. | ✔ | ✔ |
| *Clustering* | Supports clustering | Data sets can be clustered with integrated clustering approaches and varying parameters. | ✔ | ✔ |
| | Several similarity functions for pairwise object similarities | Multiple similarity functions are provided, for the calculation of pairwise object similarities from the input data matrix. | ✔ | ✔ |
| | Several integrated clustering methods | Multiple clustering methods are integrated, for the clustering of input data sets, and for the integrated comparison of method performances. | ✔ | ✔ |
| | Support of multiple object replicates | Multiple replicates per object are supported during clustering to improve accuracy, and aggregation of object replicates prior to the analysis is not required. This is a common |  ✔ | ✔ |

| | | | | | |
|---|---|---|---|---|---|
| | | scenario in life sciences when experiments are carried out several times. | | | |
| | Feature weights | Feature weights are respected during the clustering process, i.e. features contribute to the assessment of object-cluster memberships to varying degrees. | | ✔ | ✗ |
| *Clustering Statistics* | P-values for clusters | P-values and measures of statistical significance for identified clusters are provided, that allow an integrated analysis of the provided clusters from a statistical perspective. | | ✔ | ✔ |
| *Human Augmented Clustering* | Add user-defined clusters | The user can manually add clusters to the clustering with user-defined prototypes. These may be known to be relevant in the field, but may not emerge from the clustering process. | | ✔ | ✔ |
| | Modify clusters, e.g. remove, merge, or split clusters | The user can modify the clusters of the clustering by removing objects, merging multiple clusters together, or splitting a cluster into several smaller clusters. | | ✔ | ✔ |
| | Modify object-cluster assignment, e.g. delete objects, remove least-fitting objects | The user can influence the assignment of objects to clusters, and also remove objects from the clustering completely. | | ✔ | ✔ |
| *Visualizations* | Cluster visualizations | Visualizations of clusters are provided, that allow for the easy interpretation and quality control of results. | | ✔ | ✔ |
| | Network visualization of clusters | Clusters can be visualized in the context of a provided network, for instance by coloring network nodes according to their cluster membership. | | ✔ | ✔ |

|  | Cluster-network connectivity | Network connectivities of clusters can be visualized, to allow for the easy identification of enriched or depleted number of interactions between clusters on the network level. | ✔ | ✗ |
|---|---|---|---|---|
| *Reproducibility* | Operations log for reproducible workflow | A log of all performed operations is kept, to allow for the reproduction of the results. | ✔ | ✔ |
|  | Export of results for publication | Results can be exported into a format that can be used for publication. | ✔ | ✗ |
| *Network Analysis of Clusters* | Map clusters to network | Objects of the clusters can be mapped to nodes in the network, to allow network operations based on the clusters. | ✔ | ✔ |
|  | De-novo network enrichment of clusters | De-novo network enrichment approaches are integrated, to allow identification of subnetworks in a given network that are enriched in one or multiple identified clusters. | ✔ | ✔ |
|  | Connectivity analysis of clusters | Connectivities of clusters on a provided network can be assessed, and observed as well as expected connectivities can be compared with each other. | ✔ | ✗ |
|  | P-values of cluster connectivity | Connectivity p-values and measures of statistical significance for the assessed connectivity of clusters in the network are provided. | ✔ | ✗ |
| *Comparing multiple Clusterings* | Differential analysis of clusters of Two phenotypes | Two clusterings of the same objects under different experimental conditions (i.e. phenotype) can be compared, to identify clusters with differential prototype, but similar object members. | ✔ | ✗ |

| Other | Open Source | The source code of the application is publicly available. | ✔ | ✔ |

## *Supplementary Table 2: Related Work*

Related work of TiCoNE with sufficient overlap in provided functionality.

| *Tool* | *Description* | *Comparison with TiCoNE* |
|---|---|---|
| *Clustrophile* | Clustrophile (Demiralp, 2016) is an iterative clustering method, with main focus on interactive dimensionality reduction of the data features, quick iteration over different parameters, as well as comparison of multiple clustering instances. | • No human-augmented clustering<br>• No network integration of derived clusters<br>• No support for multi-replicate samples |
| *NeAT* | NeAT (Brohée, et al., 2008) is a collection of programs that allow for basic network as well as cluster analyses, also in an integrated manner. Clusters can be compared with each other to identify differential cluster members. | • No clustering functionality; only import of clusterings supported<br>• No support for multi-replicate samples |
| *NEAT* | NEAT (Signorelli, et al., 2016) is a network analysis test that allows for the statistical evaluation of connectivities of parts of a given network. It is applicable to clusters in the sense, that cluster-induced sub networks can be analyzed for enriched or depleted number of interactions between them. | • No clustering functionality<br>• No visualizations |
| *STEM* | STEM (Ernst and Bar-Joseph, 2006) is a clustering method for time-series data with few time-points, such as gene expression profiles. | • No human-augmented clustering<br>• No network integration of derived clusters<br>• No support for multi-replicate samples. |
| *VANTED* | VANTED (Rohn, et al., 2012) is a systems biology framework, tackling a larger set of common tasks. This includes data visualization, and basic cluster analysis using Self Organizing Maps. | • No human-augmented clustering<br>• No support for multi-replicate samples |
| *VISTA* | VISTA (Keke and Ling, 2003) is a visual framework that can visualize results of any clustering algorithm and allows the user to influence the clustering process based on domain knowledge. | • No network integration of derived clusters<br>• No support for multi-replicate samples |

## Supplementary Table 3: Feature Comparison with Related Work

Support of the main features of TiCoNE in related work.

|  | Clustrophile | NeAT | NEAT | STEM | VANTED | VISTA | TiCoNE |
|---|---|---|---|---|---|---|---|
| *Data Handling* | | | | | | | |
| Feature Scaling, feature selection/weights | ✔ | ✗ | ✗ | ✗ | ✔ | ✗ | ✔ |
| Removing objects with low agreement between replicates | ✗ | ✗ | ✗ | ✗ | ✗ | ✗ | ✔ |
| *Clustering* | | | | | | | |
| Supports clustering | ✔ | (✔) [a] | ✗ | ✔ | ✔ | ✔ | ✔ |
| Several similarity functions for pairwise object similarity calculation | ✔ | ✗ | ✗ | ✗ | ✗ | ✗ | ✔ |
| Several integrated clustering methods | ✔ | (✔) [b] | ✗ | ✔ | ✗ | ✗ | ✔ |
| Support of multiple object replicates | ✗ | ✗ | ✗ | ✗ | ✗ | ✗ | ✔ |
| Feature weights, i.e. time-points | (✔) [c] | ✗ | ✗ | ✗ | ✗ | ✗ | ✔ |
| P-values for clusters | ✗ | ✗ | ✗ | ✔ | ✗ | ✗ | ✔ |
| *Human Augmented Clustering* | | | | | | | |
| Add user-defined clusters | ✗ | ✗ | ✗ | ✗ | ✗ | ✗ | ✔ |
| Modify clusters, e.g. remove, merge, or split clusters | ✗ | ✗ | ✗ | ✗ | ✗ | (✔) [d] | ✔ |
| Modify object-cluster assignment, e.g. delete objects, remove least-fitting objects | ✗ | ✗ | ✗ | ✗ | ✗ | ✗ | ✔ |
| *Visualizations* | | | | | | | |
| Cluster visualizations | ✔ | ✗ | ✗ | ✔ | ✔ | ✔ | ✔ |

---

[a] Only import of clusterings supported
[b] Only graph clustering methods included
[c] Only feature selection, i.e. weights $w \in \{0,1\}$
[d] Only merge and split of clusters

| | | | | | | | |
|---|---|---|---|---|---|---|---|
| Network visualization of clusters | ✗ | ✔ | ✗ | ✗ | ✔ | ✗ | ✔ |
| Cluster-network connectivity | ✗ | ✗ | ✗ | ✗ | ✗ | ✗ | ✔ |
| *Reproducibility* | | | | | | | |
| Operations log for reproducible workflow | ✗ | ✗ | ✗ | ✗ | ✗ | ✗ | ✔ |
| Export of results for publication | ✗ | ✗ | ✗ | ✗ | ✗ | ✗ | ✔ |
| *Network Analysis of Clusters* | | | | | | | |
| Map clusters to network | ✗ | ✔ | ✔ | ✗ | ✔ | ✗ | ✔ |
| De-novo network enrichment of clusters | ✗ | ✗ | ✗ | ✗ | ✗ | ✗ | ✔ |
| Connectivity analysis of clusters | ✗ | ✗ | ✔ | ✗ | ✗ | ✗ | ✔ |
| P-values of cluster connectivity | ✗ | ✗ | ✔ | ✗ | ✗ | ✗ | ✔ |
| *Comparing multiple clusterings* | | | | | | | |
| Differential analysis of clusters of two phenotypes | ✗ | ✔ | ✗ | ✔ | ✗ | ✗ | ✔ |
| *Other* | | | | | | | |
| Open Source | ✗ | ✗ | ✔ | ✔ | ✔ | ✗ | ✔ |

**Supplementary Figure 1. Artificial illustration data.** Temporal prototypes P1-P3 of the artificial time series clusters CL1-CL3 and their in-between edge probabilities used to generate an artificial network. Random objects (RND) do not show a common temporal profile. Cluster sizes are indicated in brackets. Edge labels correspond to undirected edge probabilities in the synthetic network. We constructed the network with enriched numbers of edges between the cluster pairs connected by orange arcs in the figure, and with depleted edge numbers if connected by blue arcs. Using this toy example, we explain how to extract "*de novo*" (1) the three temporal prototypes P1-P3, and (2) the aforementioned network connectivity enrichments and depletions.

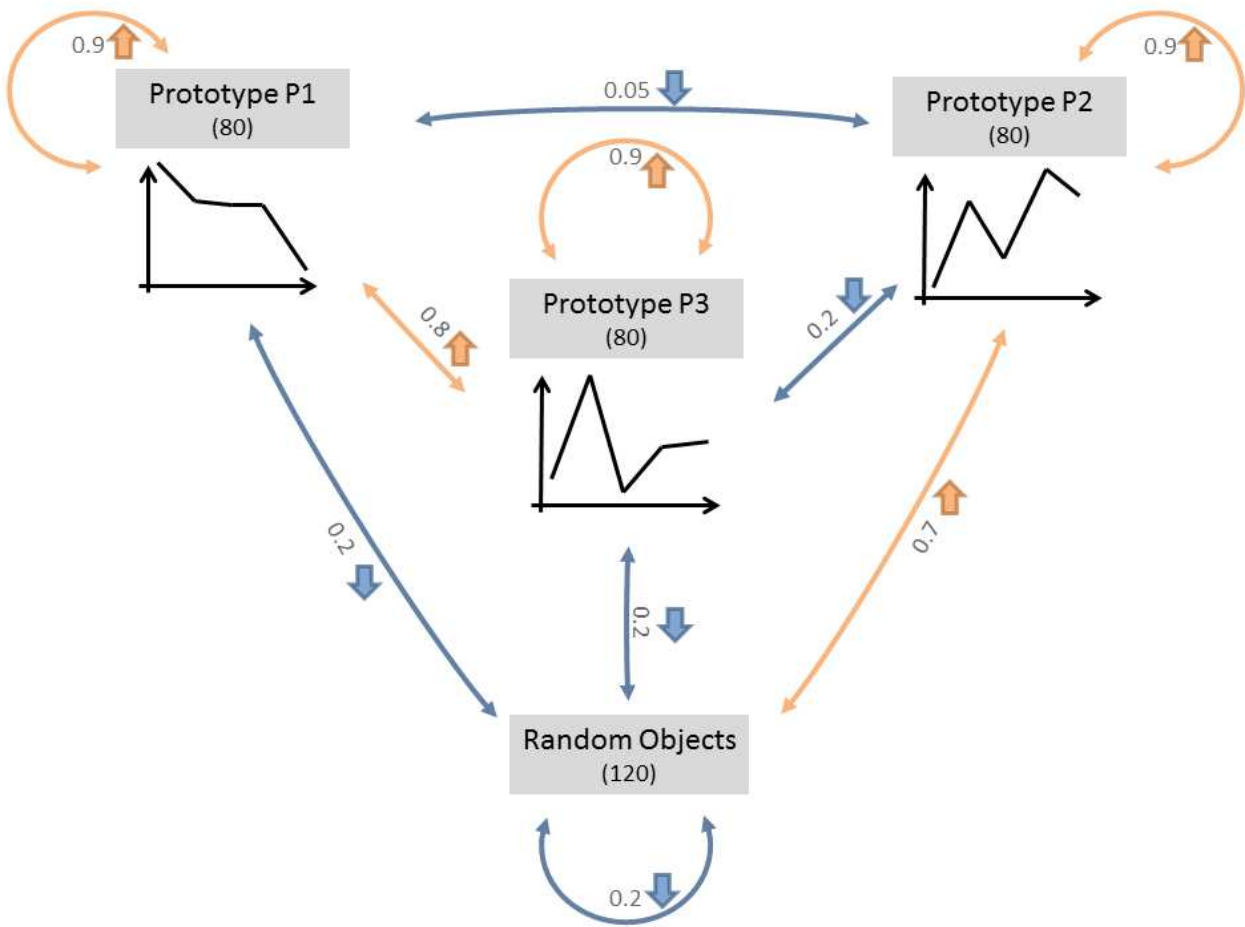

**Supplementary Figure 2**. **Performance illustration on artificial data.** **(a)** Clustering of the synthetic data set into eight clusters. Clusters 5, 11 and 9 correspond to artificially generated prototypes 1, 2, and 3, respectively (compare Supplementary Figure 1). Clusters are sorted by p-value and insignificant clusters ($p \geq 0.01$) are shaded. **(b)** Co-enrichment of all eight clusters reveals the cross-connectivity artificially hidden in the network (see Supplementary Figure 1). Only significant enrichments and depletions ($p \leq 0.01$) are shown.

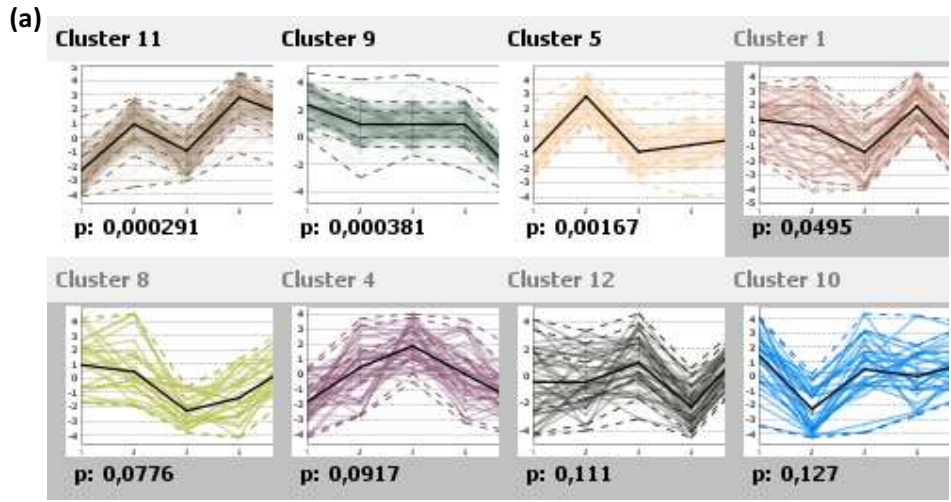

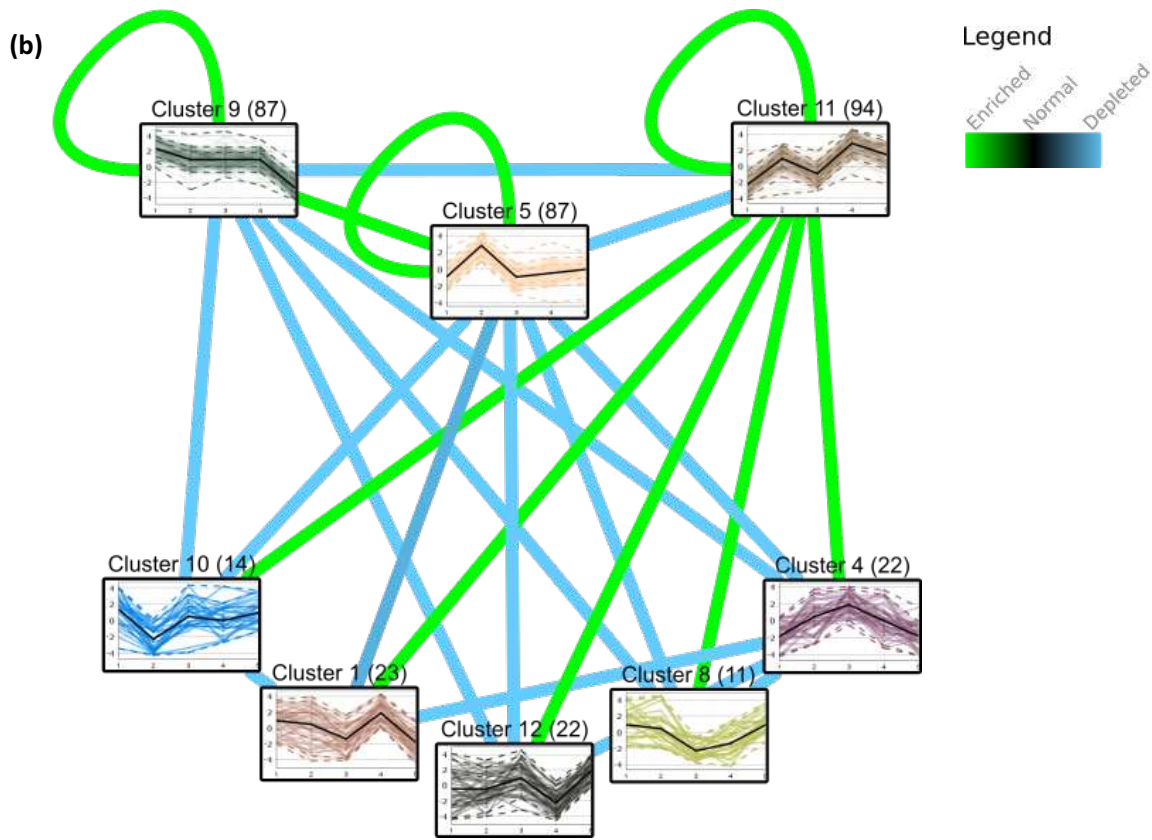

**Supplementary Figure 3**. **Temporal gene expression patterns after Influenza and Rhino virus infection.** Significant clusters ($p \leq 0.05$) of the final clustering of the **(a)** Influenza and **(b)** Rhino data set. Clusters are sorted by prototype standard variance. Note the individual y-axis scaling.

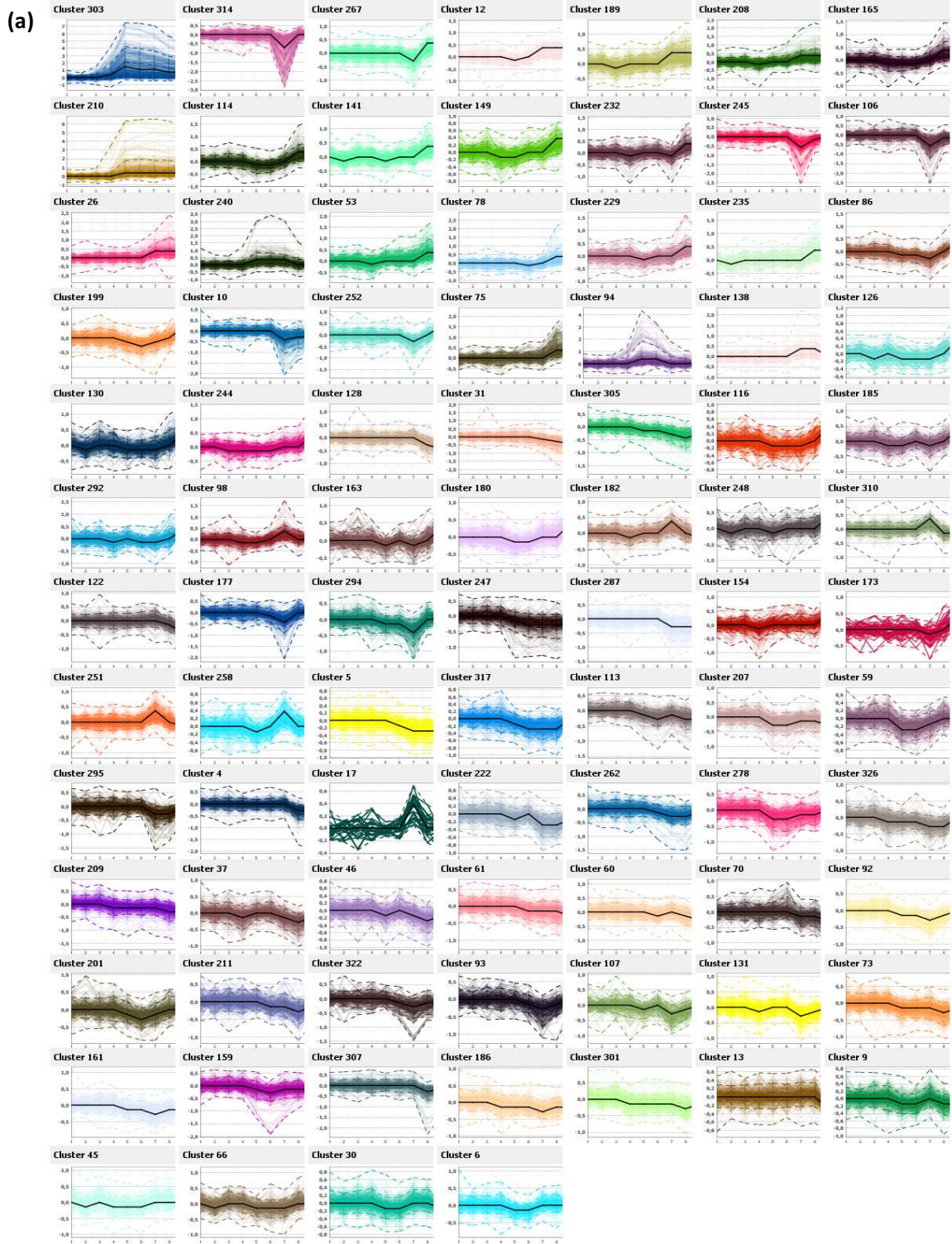

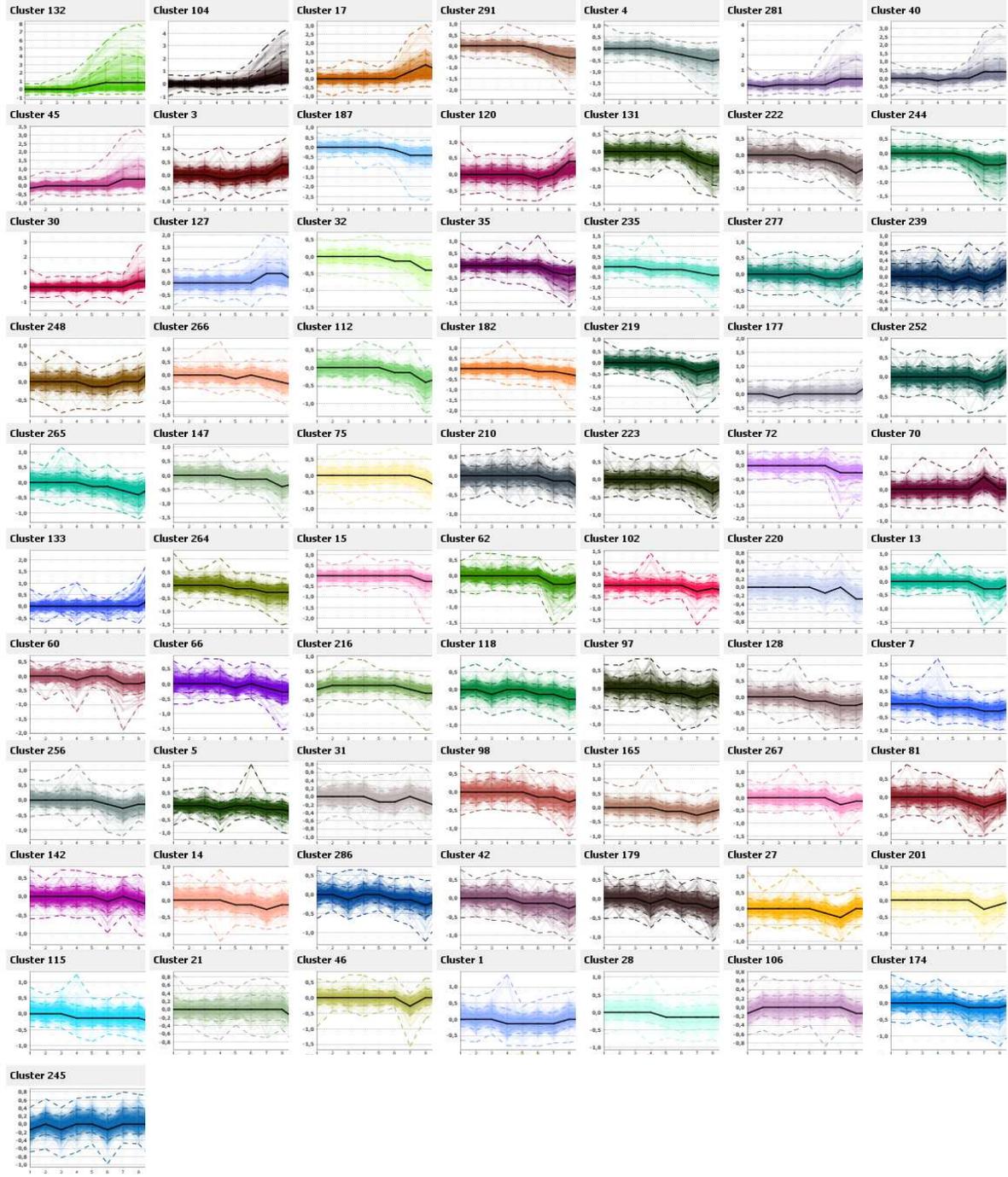

**Supplementary Figure 4**. **Influenza time course network co-enrichment.** (a) Co-enrichment representation of temporal expression patterns after Influenza infection expressed by genes that interact more (less) often than expected by chance. The edge colors correspond to log-odds scores. Only edges with $p \leq 0.02$ are shown. Nodes are labeled with the corresponding cluster number and their size (number of objects) in brackets. Clusters are arranged along a timeline regarding their peak variance. (b) Top 5 most significant co-enrichment cluster pairs.

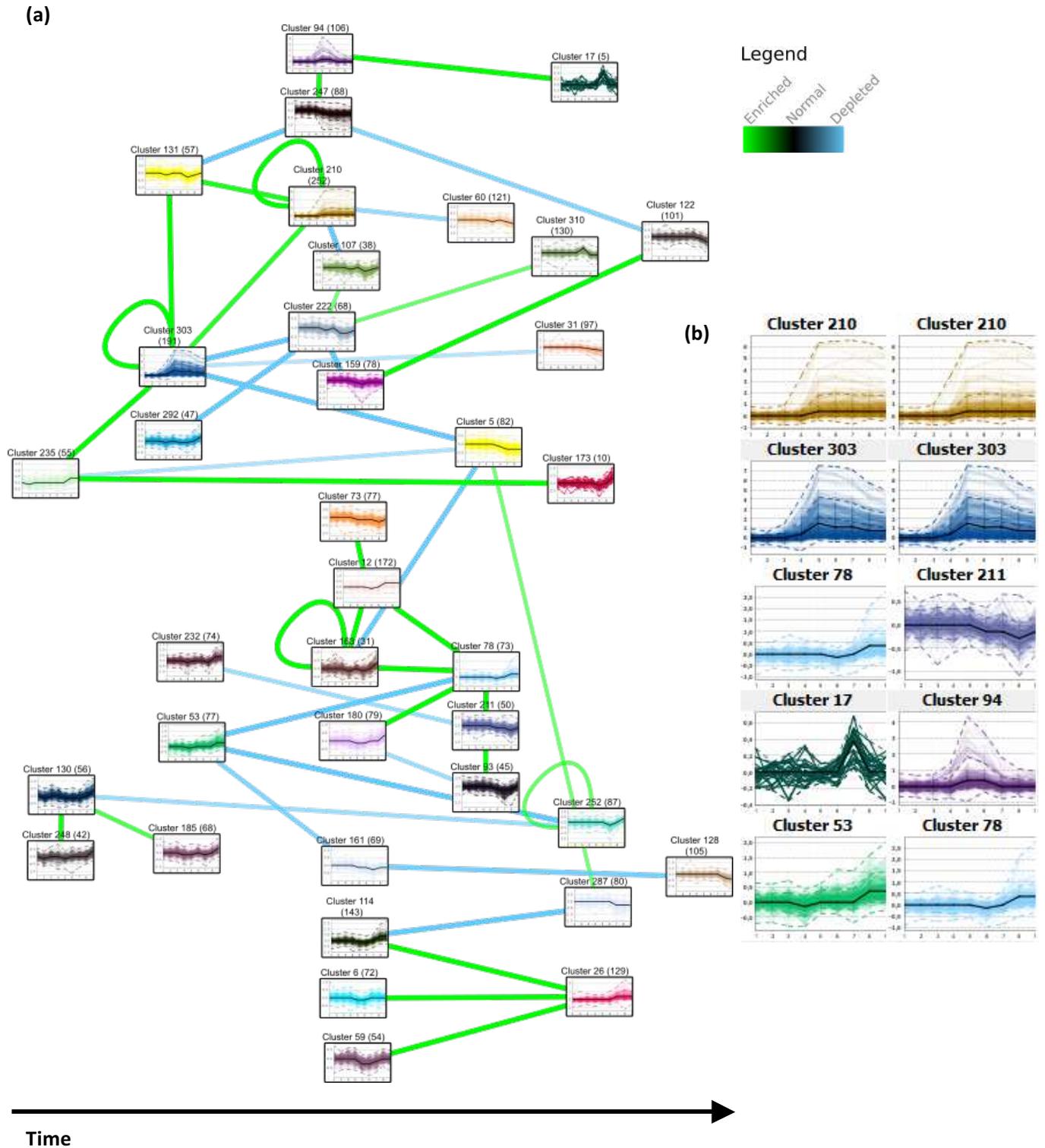

**Supplementary Figure 5**. **Rhino virus time course network co-enrichment. (a and b)** Co-enrichment representation of temporal expression patterns after Rhino virus infection; analogous to Supplementary Figure 4.

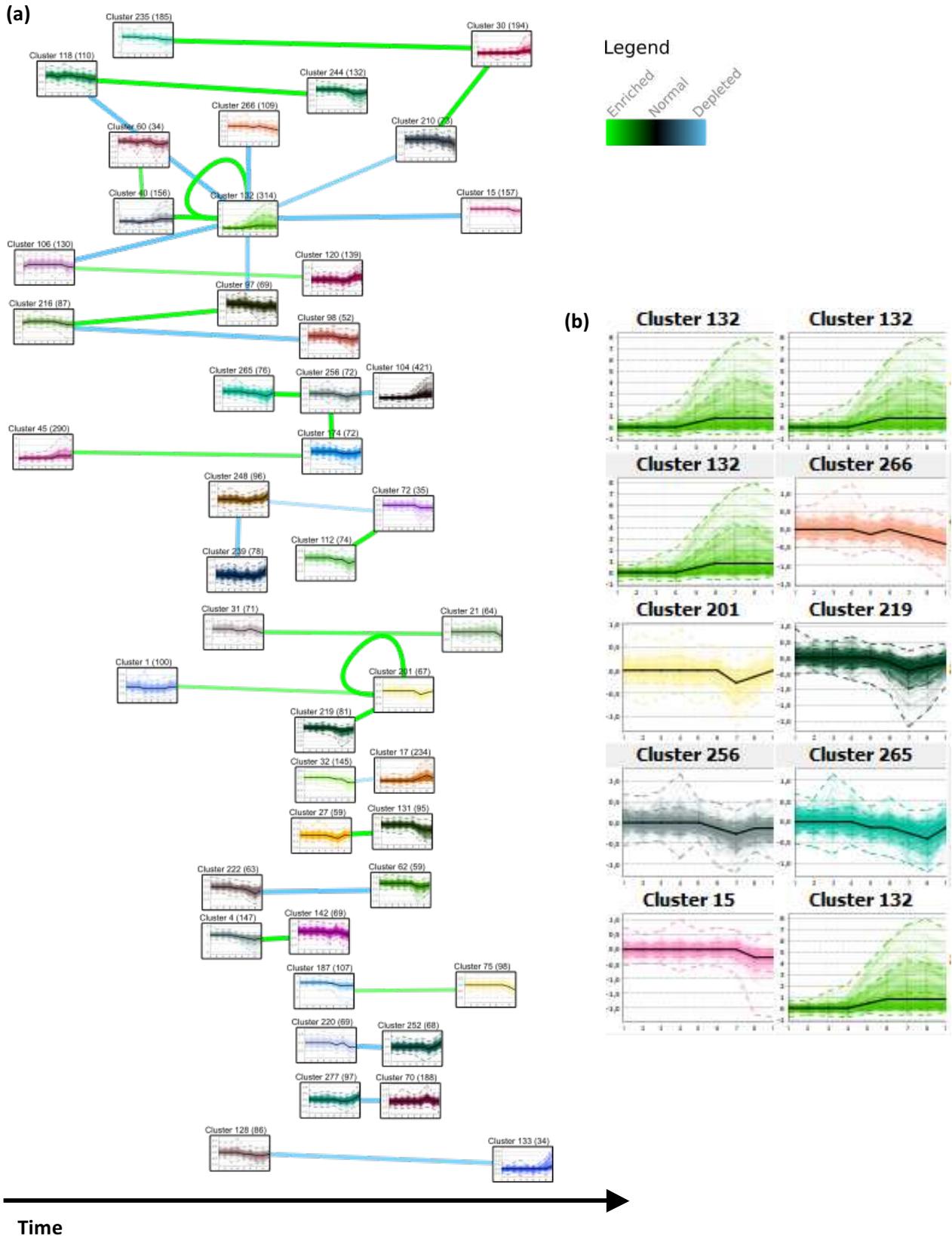

**Supplementary Figure 6**. Largest connected network component after phenotype comparison of the Influenza and Rhino virus experiments showing all significant cluster pairs ($p \leq 0.01$, see text).

**Supplementary Figure 7.** Histograms of edge counts between cluster pairs after co-enrichment analysis for **(a)** Influenza data, and **(b)** Rhino data. Histograms are truncated with the same value to enhance comparability.

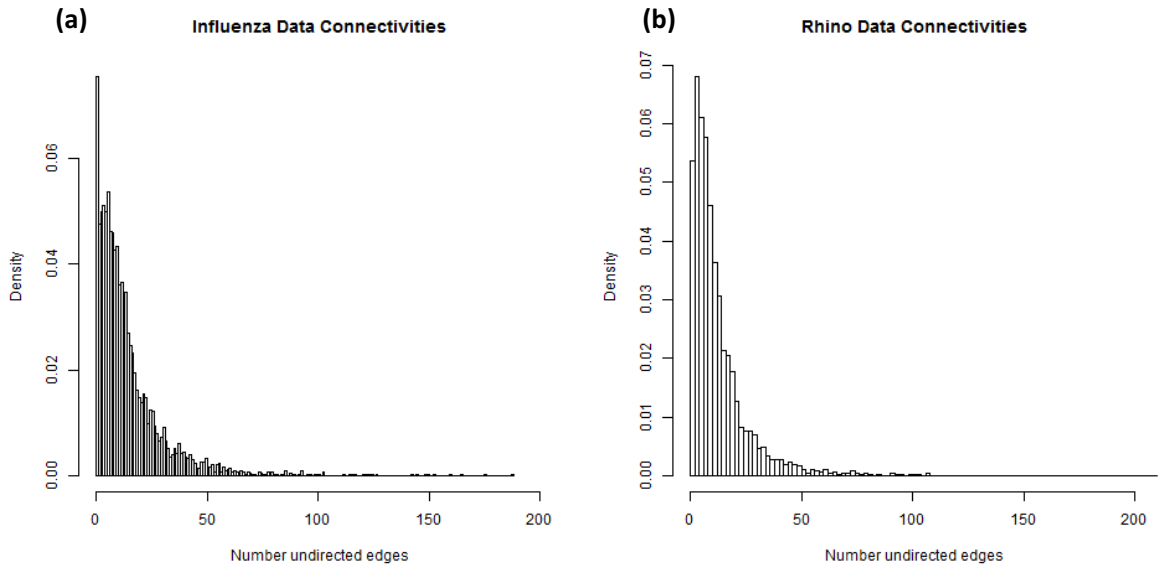

**Supplementary Figure 8.** Histograms of log-odds scores between cluster pairs after co-enrichment analysis for **(a)** Influenza data, and **(b)** Rhino data.

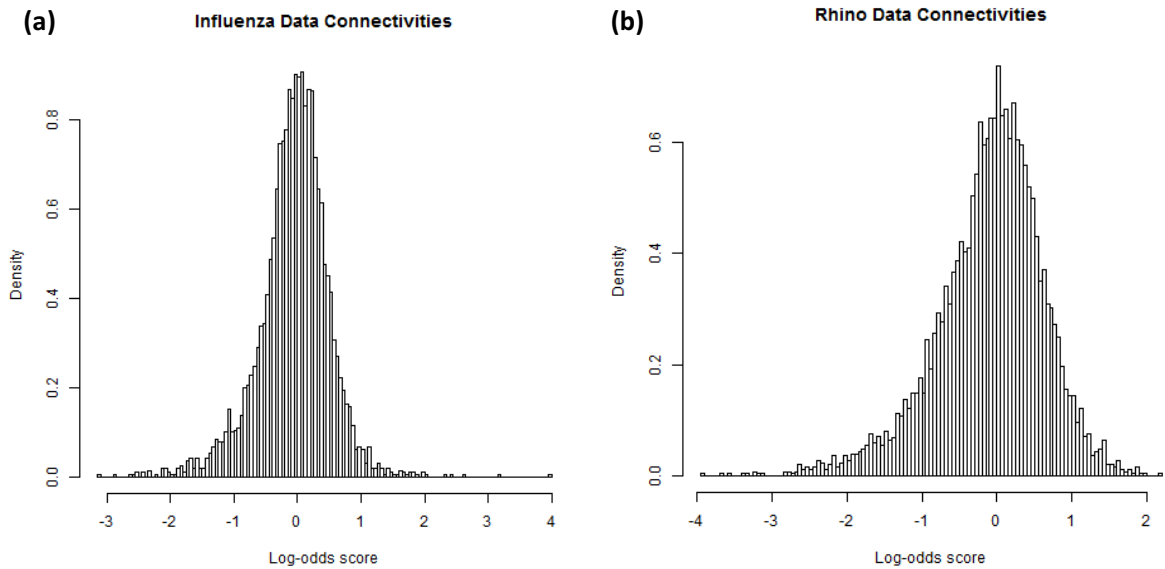

**Supplementary Figure 9.** Scatter plots of p-values versus log-odds scores for **(a)** Influenza data, and **(b)** Rhino data after co-enrichment analysis.

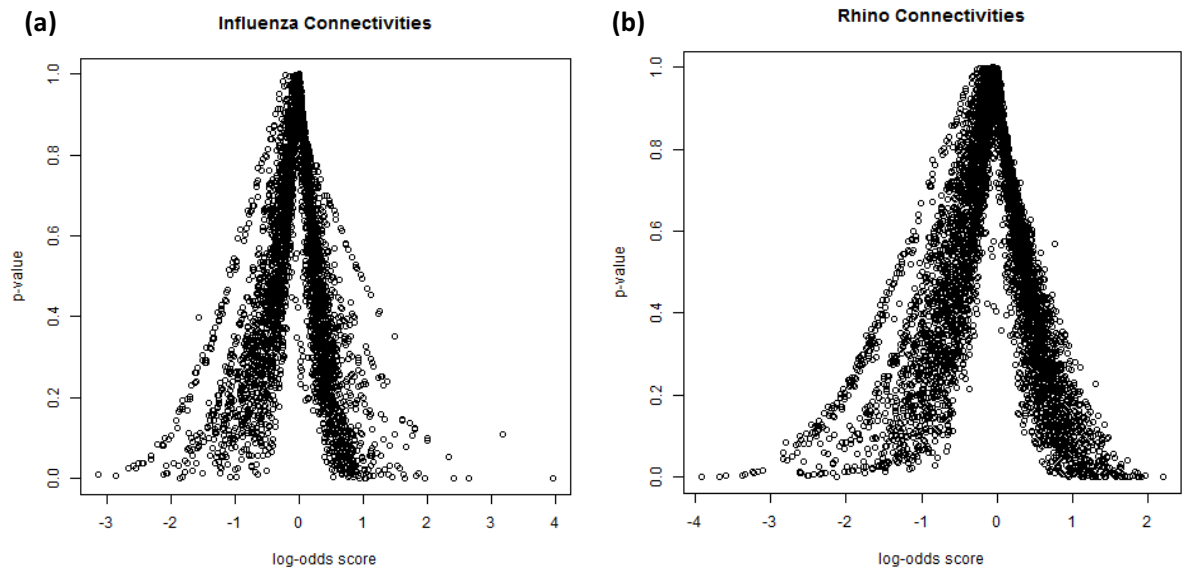

**Supplementary Table 1**. **Properties of the eight clusters found in the artificial data.** Clusters are ranked by p-value. Average gene-to-prototype similarity in terms of the average Pearson correlation of genes to their respective cluster prototype and the corresponding p-value. A visualization of this table is given in Supplementary Figure 2a.

| Cluster | # Objects | Average gene-prototype Pearson correlation | p-value |
|---|---|---|---|
| **11** | **94** | **0.937** | **0.000291** |
| **9** | **87** | **0.939** | **0.000381** |
| **5** | **87** | **0.918** | **0.00167** |
| 1 | 23 | 0.892 | 0.0495 |
| 8 | 11 | 0.891 | 0.0776 |
| 4 | 22 | 0.878 | 0.0917 |
| 12 | 22 | 0.873 | 0.111 |
| 10 | 14 | 0.876 | 0.127 |